\documentclass[10pt,twocolumn]{article}
\usepackage{amsmath}
\usepackage{multirow}
\usepackage{graphicx}
\usepackage{subcaption}
\usepackage[margin=2cm]{geometry}
\usepackage{cite}
\usepackage{etoolbox}
\usepackage{subfig}
\usepackage{adjustbox}
\usepackage{array}
\usepackage{authblk}

\title{The Impact of Surface Geometry, Cavitation, and Condensation on Wetting Transitions: Posts and Reentrant Structures}
\author[1]{Panter J. R.}
\author[1]{Kusumaatmaja H.}
\affil[1]{Department of Physics, Durham University, South Road, Durham, DH1 3LE}

\date{}
\begin{document}
\twocolumn[
\begin{@twocolumnfalse}
\maketitle

\begin{abstract}
The fundamental impacts of surface geometry on the stability of wetting states, and the transitions between them are elucidated for posts and reentrant structures in both two and three dimensions. We identify three principal outcomes of particular importance for future surface design of liquid-repellent surfaces. Firstly, we demonstrate and quantify how capillary condensation and vapour cavitation affect wetting state stabilities, and the roles condensates play in wetting transitions. Crucially, this leads to a description of the surface structures which exhibit a suspended state in the absence of a collapsed state. Secondly, two distinct collapse mechanisms are observed for 3D reentrant geometries, Base Contact and Pillar Contact, which are operative at different pillar heights. As well as morphological differences in the penetrating liquid, each mechanism is affected differently by changes in the contact angle with the solid. Finally, symmetry breaking is shown to be prevalent in 2D systems, but absent in the 3D equivalents. For the 2D reentrant geometries, three pillar heigh-dependent collapse mechanisms are shown: asymmetric Pillar Contact and Base Contact, and a third hybrid mode.

\end{abstract}
\end{@twocolumnfalse}
]

\section{Introduction}
Superamphiphobic surfaces demonstrate contact angles in excess of $150^o$ and low contact angle hysteresis (less that $10^o$) for liquids with low surface tensions, such as oils (superoleophobicity), and those with high surface tensions, for example water (superhydrophobicity) \cite{Chu2014}. Such surfaces are expected to play vital roles in engineering a sustainable and energy-efficient future by, for example, reducing fuel waste in tanks and lines \cite{Brown2016, Nosonovsky2009}, reducing biofouling and drag in marine shipping \cite{Quere2008,Genzer2006}, enabling self-cleaning \cite{Furstner2005} and improving the capabilities and performance of microfluidic devices \cite{Rothstein2010}. Fundamentally, the success of these applications lies in the ability of physically structured surface textures to maintain a suspended state (in which the liquid sits atop a composite solid-vapour surface) relative to the collapsed state (the liquid fills the texture). 

Despite the imminent need for these surfaces and the significant synthetic advances across a broad range of wetting applications \cite{Wang2015}, two key challenges remain as significant barriers to industrial development. 

The first challenge is the maintenance of the suspended state when the texture is entirely submerged in the liquid phase. Although from the analysis of interfacial tensions, an underwater suspended state was suggested to be possible \cite{Marmur2006}, solvation of the vapour phase in the liquid has meant that many experimental submerged suspended states have only demonstrated a finite lifetime, typically on the time scale of days \cite{Jones2015, Poetes2010,Bobji2009,Emami2013}. Design criteria have been suggested for increasing this lifetime, such as by utilising nanoscale \cite{Jones2015} and multi-scale textures\cite{Papadopoulos2016}, or optimising microtexture dimensions \cite{Jones2015, Xu2014}. Overwhelmingly however, experimental observations suggest that the suspended state is prone to collapse when triggered by vibrations to the system. The situation is worsened further by considering that although nanoscale textures increase the barrier to the wetting transition, large vapour pockets are required in order to maximise the slip length and increase the efficiency of driving a fluid across the texture \cite{Quere2008}.  

The second challenge is that current superamphiphobic surfaces have limited mechanical strength, engendered by the typical overhang structures, and low volume fractions of solid to air \cite{Dyett2014}. A large number of synthetic procedures have attempted to address this issue (see for example \cite{Lee2013,Hu2016,Kang2012,Kim2016,Tuteja2007} and the review \cite{Hensel2016}). However, the geometrical features which produce a robust suspended state are relatively unknown, such that it is difficult to produce a texture which is resistant to impact and abrasion without compromising the stability of the suspended state. Fundamentally the barrier to the wetting transition remains poorly elucidated for all but the simplest structures \cite{Bormashenko2015}.  

Few experimental procedures have been able to track the liquid-vapour interface throughout the collapse transition. For liquid droplets on square arrays, the transition was shown to occur over a time-scale of less than 1 ms \cite{Reyssat2008}, making the time-resolution of the interface challenging. Greater success however was achieved through following the liquid-vapour interface of the suspended state in submerged well and ridge systems \cite{Rathgen2010, Lv2014,Lv2015}. As the transition occurred over several minutes due to the rate-limiting solvation of air in water, this enabled the effective interface visualisation using of laser confocal microscopy and diffraction measurements. The principal result from these studies was that for each submerged well and ridge, the transitioning liquid-vapour interface preserved the local symmetry whilst sliding down the feature. Upon contacting the base of the system, the transition completed rapidly.

Wetting transitions on more complex structures have been able to be simulated over a broad range of length scales by utilising different computational techniques; molecular dynamics at the nanoscale, diffuse interface models for mesoscale wetting, and sharp interface models at the macroscale. Despite the significant insights offered by these studies, the precise influence of the surface geometry on the wetting mechanisms remain absent in current computational investigations. 

For example, in systems comprised of multiple posts, the wetting transition mechanism is consensually characterised as being initiated through the local collapse about a single feature, followed by lateral propagation of the liquid to fill the texture \cite{Ren2014,Zhang2014,Savoy2012,Savoy2012_2}. Predominantly, the transition state was shown to be associated with the first local collapse. The transition barrier was greater in energy for reentrant geometries compared to posts, as the reentrant geometry imposed a larger liquid-vapour interface area at the transition state.  The influence of multiple features on the wetting transition was further shown to be enhanced when the liquid droplet size was of the order of the scale of the surface structures \cite{Shahraz2014,Wang2015, Gross2010}. In the presence of multiple posts therefore, the complex local collapse - propagation mechanism means that the roles of each structural aspect of the surface geometry in the wetting transition are difficult to probe individually. In two dimensions, or with 3D ridge geometries, as the lateral propagation stage is not possible, the global collapse transition has been shown to occur through a series of local transitions \cite{Connington2013,Pashos2015a}. However, because in these works a liquid droplet was simulated, each local collapse was influenced by the droplet shape, and not by the surface geometry alone.

Attempting to eliminate the effects of multiple posts, only a small number of studies have sought to examine the structure's geometrical influence on the wetting transition. Evaluation of the Young-Laplace equation (in a sharp interface model) has previously enabled the comparison between the collapse mechanisms of reentrant and post geometries \cite{Pashos2016}. The small height to width ratio of the both structures meant that the transition proceeded by the interface sagging to contact the base of the system (the transition state) before depinning from the top of the structure. Thus, although the geometry affected the shape of the interface late in the transition, it had little bearing close to the transition state. In contrast to this, a diffuse interface model of a droplet atop a single pillar suggested that the transition was dominated by cavity condensation (which in turn is highly geometry - dependent) \cite{Pashos2015}, although this formalisation was unable to track the transition.

In two dimensions, in direct contrast to three dimensional experiments, the collapse mechanism on posts and reentrant geometries has always been observed to break the symmetry of the system \cite{Giacomello2012,Giacomello2012a, Giacomello2015, Amabili2016}. Principally, these authors observed multiple transition paths for the reentrant geometries. For the transition of a droplet on multiple 3D ridge structures, each local collapse either preserved or broke the local symmetry \cite{Pashos2015a}, although the reason for the selection of a specific mechanism was not presented. Symmetry breaking was also observed using molecular dynamics simulations on 3D post systems \cite{Prakash2016}. Here, following the contact of the liquid at the base of the system, the resulting vapour cavities collapsed asymmetrically. 

It has been our aim therefore to elucidate the geometrical influence of the surface structure on the wetting states and collapse mechanism through minimising he extraneous effects present in previous works. By using a diffuse interface model (allowing for the facile simulation of topological changes), assumptions about the liquid vapour interface morphology are negated. The minimum energy pathways, MEPs, (the steepest-descent pathways with the smallest transition state energies) obtained between the collapsed and suspended states provide a lower bound to the transition state energy obtained via any experimentally realisable transition pathway.

We compare square posts and reentrant structures in two and three dimensions in order to elucidate the impact of the reentrant cap structure. Importantly, we find the geometry to critically influence three principal aspects of wetting. Firstly, we highlight the significant impact of liquid condensation and vapour cavitation on the wetting state stabilities and transitions mechanisms (Section \ref{sec:concav}). Secondly, two distinct mechanisms are observed for the wetting transition of 3D reentrant geometries. Pillar height is shown to critically influence which mechanism is operative. The influence of  surface wettability and pressure are investigated for each (Section \ref{sec:reen}). Finally, we discuss the influence of the surface geometry on the symmetry of the transitions, finding significant symmetry breaking only in two dimensions. In 2D, three pillar height-dependent mechanisms are identified (Section \ref{sec:sym}).

\section{Methods} \label{sec:methods}
\subsection{Phase field model}
Here, to describe a bi-fluidic system in contact with a solid surface, the energy functional employed is based on a Landau-type free energy model \cite{Briant2004,Briant2004_2,Kusumaatmaja2010,Kusumaatmaja2015}. We choose the scalar order parameter $\phi(\textbf{r})$, representing the local composition, such that $\phi=1$ in the pure liquid phase and $\phi=-1$ in the pure vapour phase. According to Cahn \cite{Cahn1977}, treating the fluid-surface interactions as short range enables us to decompose the system free energy $\Psi$, into
\begin{equation}
\Psi-\Psi_o = \Psi_i + \Psi_s - \Delta P V_l. \label{eqn:psi}
\end{equation}
Here, $\Psi_i$ is the free energy contribution arising from an isotropic biphasic system \cite{Cahn1958},
\begin{equation}
\Psi_i =\int_V \left(\psi_b + \frac{\epsilon}{2} \vert \nabla \phi \vert^2 \right)dV,
\end{equation}
in which the first term of the integrand is a Landau-type free energy density for a homogeneous (bulk) system $\psi_b = \frac{1}{\epsilon} \left(\frac{1}{4} \phi^4 - \frac{1}{2}\phi^2 \right)$. The second term describes the energy density associated with composition gradients and as such represents a surface tension.  

$\Psi_s$ is the free energy associated with solid-liquid interactions, approximated as an integral over the surface,
\begin{equation}
\Psi_s = \int_S \psi_s dS,
\end{equation}
in which to a first approximation the surface energy density is represented as $\psi_s = -h\phi_s$ \cite{Cahn1977,DeGennes1985}. $\phi_s$ is the value of the order parameter at the surface. Previously $h$ has been referred to as a wetting potential, and is related to the contact angle of a liquid droplet on a planar surface $\theta_o$ through
\begin{equation}
h=\rm{sign}\left( \frac{\pi}{2}-\theta_o \right) \sqrt{2\cos \left(\frac{\alpha}{3}\right)\left[1-\cos\left(\frac{\alpha}{3}\right)\right]},
\end{equation}
where $\alpha = \arccos(\sin \theta_o)$ and the function `sign' returns the sign of the argument \cite{Papatzacos2002}.

In this work, the effect of pressurising the system is introduced into this phase field model. This is incorporated into the free energy functional through the addition of the terms $-P_l V_l$ and $-P_v V_v$ in terms of the liquid and vapour pressures, $P_l$ and $P_v$, and volumes, $V_l$ and $V_v$, respectively. Within this phase field model,
\begin{equation}
V_l = \int_V \frac{\phi+1}{2} dV,
\end{equation}
and by using $V_v = V-V_l$ and $\Delta P = P_l-P_v$, with $V$ as the simulation volume, the pressure terms can be rearranged into the form $-\Delta P V_l + P_vV$. This pressure treatment is similar to that employed previously in a two dimensional, sharp interface model \cite{Amabili2016}. We label $P_vV$ as $\Psi_o$ with which we chose to reference $\Psi$ against, describing a system occupied by only the vapour phase at $P_v=0$ with all surfaces having $\theta_o = 90^o$. For the remainder of this report, $\Psi-\Psi_o$ is denoted as $\widetilde{\Psi}$. As has been previously reported, the pressure term is equivalent to a chemical potential \cite{Giacomello2012,Evans1986}. The effect of this term is to effectively contact the system with an external reservoir at constant pressure, enabling the exchange of both phases at any point within the system. 

The solid-vapour, solid-liquid and liquid-vapour surface tensions, $\gamma_{sv}, \gamma_{sl}$ and $\gamma_{lv}$ can be expressed by considering the excess bulk free energy density \textit{W} in the presence of an interface (see for example \cite{Kusumaatmaja2010}),
\begin{align}
	\gamma_{lv}&=\sqrt{\frac{8}{9}}, \label{gammalv} \\
	\gamma_{sl}&=\frac{\gamma_{lv}}{2}\left(1-(1+\sqrt{2}h)^{3/2}\right), \label{gammasl} \\
	\gamma_{sv}&=\frac{\gamma_{lv}}{2}\left(1-(1-\sqrt{2}h)^{3/2}\right). \label{gammasv}
\end{align}

In this work, in a similar manner to a previous treatment \cite{Kusumaatmaja2015}, the computational domain is discretized into a $N_x \times N_y \times N_z$ cubic lattice of points (nodes), each associated with a value $\phi_{ijk}$, where $i,j,k\in\left\lbrace1,...,N_{x,y,z}\right\rbrace$. A spatial separation between adjacent points is defined, labelled \textit{G}. These nodes are classified according to whether they are at the surface of a solid boundary (surface nodes), within the solid (solid nodes), or within the bulk system (bulk nodes). Although the solid nodes are always assigned an initial $\phi$, they do not contribute to the free energy or the free energy gradients of the system, and are not updated in the energy minimization. Furthermore, in order that the solid nodes do not impact the transition pathway, the eigenvalues corresponding to changing $\phi$ at these nodes are shifted, and the corresponding eigenvectors are projected out of the Hessian \cite{Wales2003}. Details of the discretization methodology can be found in the Supplementary Information.

\subsection{Simulation implimentation} \label{sec:simulationparameters}
\begin{figure}[!ht]
\centering
\includegraphics[width=0.45\textwidth]{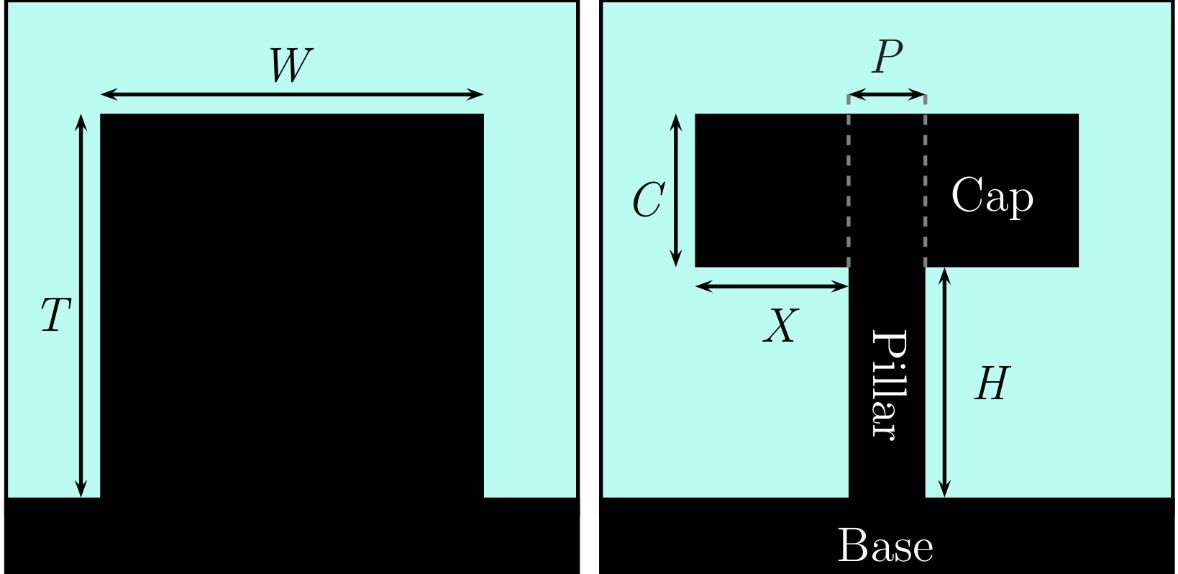}
\caption{Geometric descriptions of the post (left) and reentrant geometry (right) with dimensions labelled.}
\label{fig:geom}
\end{figure}

In the following sections, the surface geometries are termed `post' and `reentrant' depending on the cap structure. Cross sections through the geometries are shown in Fig. \ref{fig:geom}. 
Here, simulations are carried out in two and three dimensions, with the heights of the post and reentrant structures varied through changing \textit{T} and \textit{H}. Two specific heights are referred to throughout, labelled `Short' (\textit{T} = 30\textit{G}) and `Tall' (\textit{T} = 50\textit{G}). The other dimensions remain fixed throughout, such that (in units of G): \textit{W} = 29, \textit{X} = 12, \textit{C} = 12, and \textit{P} = 5. One structural replica is used per simulation system. In the 3D simulations, the dimensions are the same in the x and y directions.

The size of the structural features was chosen to balance the computational cost of the simulation with the Cahn number, $C_n$, defining the ratio of the interface width to the smallest length scale in the system. It has been suggested that $C_n \leq 0.5$ is sufficient for the diffuse interface to not influence the behaviour of the system significantly \cite{Connington2013}. To ensure this condition, here $C_n \leq 0.2$. Further confidence in this assumption is gained in the observation that, in molecular dynamics simulations, structures need only be approximately three molecules wide in order to exhibit macroscopic wetting behaviour \cite{Leroy2012}. 

For the 2D and 3D systems, $N_y$ = 60. For short geometries, $N_z$ = 50, whilst for tall geometries, $N_z=60$. In the 3D simulations, $N_x=60$. The interface width $\epsilon$ is chosen as $\epsilon=0.02$. Because the lattice must have sufficient resolution to resolve the interface, the lattice spacing $G=0.02$ is also used. Again, this was chosen in order to balance the computational cost of the simulation with the accuracy with which the interface was resolved. 

By dividing $\widetilde{\Psi}$ by $\gamma_{lv}A_f$, in which $A_f$ is the area of the base of the system,$\widetilde{\Psi}$ is nondimensionalzed, labelled $\widetilde{\Psi}_r$. In a similar manner, $\Delta P$ is divided by $\frac{\gamma_{lv}}{N_yG}$ to make $\Delta P_r$ dimensionless.

Through trialling basin hopping steps \cite{Wales2003, Wales1997, Li1987}, a complete catalogue of the system free energy minima could be obtained through initialising a planar fluid-vapour interface at heights 0, \textit{H}, \textit{T} and $N_z$. Below the interface, $\phi_{ijk}=-1$, at the interface $\phi_{ijk}=0$ and above $\phi_{ijk}=+1$. The minimization convergence criterion used was that the gradient at each node was separately converged to a tolerance of $10^{-11}$. Because of the relatively small number of nodes located at liquid-solid, liquid-vapour or solid-vapour interfaces relative to the system volume, an rms convergence criterion was found to be unsuccessful in converging to the true free energy minima.

The large simulation sizes used in this work (of the order of $10^5$-$10^6$ degrees of freedom) necessitated the use of a memory-efficient minimization method, here we use the limited-memory Broyden-Fletcher-Goldfarb-Shanno (LBFGS) algorithm \cite{Nocedal1980,Liu1989}. In order to compute both the transition states between minima, and the transformations between them along the MEPs, the doubly-nudged elastic band (DNEB) method is used \cite{Henkelman2000,Trygubenko2004} within the program OPTIM \cite{OPTIM}. 

The initial DNEB string was implemented with 37 images (inclusive of the minimum end-points). As the MEP converged upon by DNEB is sensitive to the initial string, several strings were trialled for the 2D and 3D systems; either preserving or breaking the system symmetry. This distinction was however found to be important only in the case of the 2D post geometries.

\section{Condensation and Cavitation} \label{sec:concav}
Condensation of the liquid from the vapour phase, and cavitation of the vapour from the liquid phase, are found to critically influence the stabilities of the free energy minima. Within this section, we quantify these effects and principally describe the stability limits of each minimum in $\theta_o$ and $\Delta P_r$. We also show that condensation plays a key role in determining the minimum energy pathways for the collapse mechanism of fluids with low $\theta_o$. Overall, we elucidate how each structural aspect of the surface geometry impacts the formation of condensates along the minimum energy pathway.

In order to describe the wetting characteristics of the surface textures in terms of common experimentally controlled parameters, we first survey the free energy minima for the contact angle range $0^o \leq \theta_o \leq 180^o$, and the pressure range $-0.25 \leq \Delta P_r \leq 0.25$. The energy minima in 2D are discussed first.
   
\begin{figure*}[!ht]
\centering
\includegraphics[width=\textwidth]{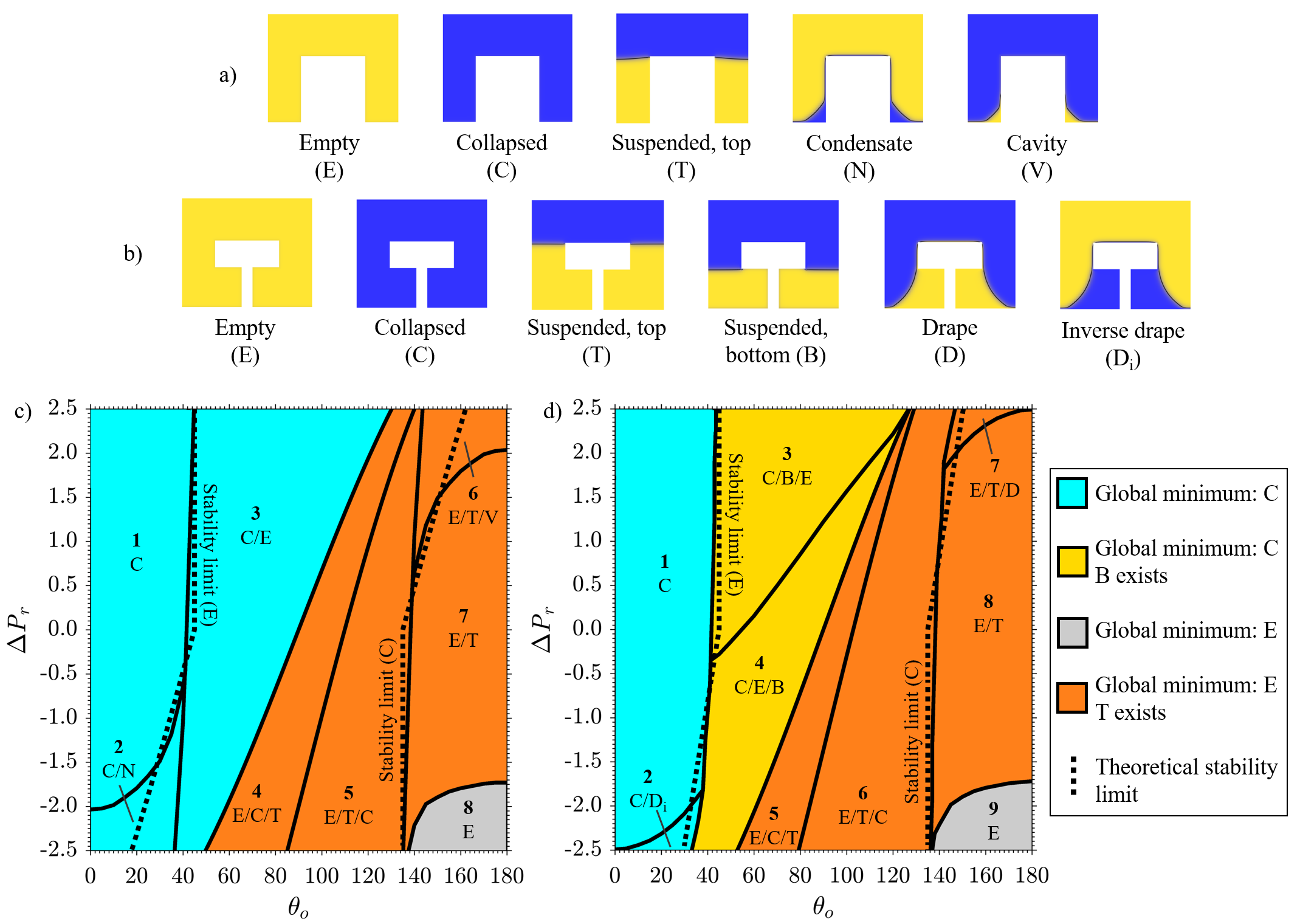}
\caption{Images of the phase field at the free energy minima for 2D posts (a) and reentrant geometries (b). The solid structure is shown in white, the liquid phase ($\phi=1$) in blue, and the vapour phase ($\phi =-1$) in yellow. The $\phi=0$ boundary line is highlighted in black. The phase diagrams are shown for the post (c) and reentrant geometry (d), with numbered regions indicating the stable states listed in order of increasing energy.}
\label{fig:2dpostmin}
\end{figure*}

For the 2D posts, six minima are identified and imaged in Fig. \ref{fig:2dpostmin} (a). In addition to the vapour-filled (Empty), liquid-filled (Collapsed) and suspended minima, the product of heterogeneous nucleation of the liquid phase in the empty state (Condensate) is observed, with a third evidencing the nucleation of vapour within the liquid-filled system (Cavity).

For the 2D reentrant geometries, seven minima are highlighted and shown in Fig. \ref{fig:2dpostmin} (b). As with the posts, the reentrant geometry also admits the empty, suspended (top) and collapsed solutions. Here however, for $\theta_o < 90^o$, the suspended (top) state becomes unstable with respect to the liquid-vapour interface sliding down the cap, leading to a different suspended state in which the interface is pinned at the base of the cap (suspended, bottom). This latter minimum highlights the ability of the reentrant structure to suspend liquids which would otherwise wet an unstructured solid substrate. Condensates and cavities are observed to form at the base of the pillar and under the cap. However, these solutions are only stable over a narrow $\theta_o-\Delta P_r$ coordinate range as the condensates and cavities readily coalesce to fill the underside of the reentrant cap structure, leading to the Drape and Inverse Drape states.

Evaluation of the existence and energy of each minimum type across the range of $\theta_o-\Delta P_r$ coordinates enables a phase diagram to be constructed for the post and reentrant geometries, shown in Figs. \ref{fig:2dpostmin} (c) and (d) respectively. Initially, we focus on the three phase diagram features in the approximate range $45^o < \theta_o < 135^o$, which are not strongly influenced by cavitation and condensation effects: the boundary at which the suspended (top) and collapsed states become isoenergetic, the boundary at which the suspended (bottom) and empty states become isoenergetic (for the reentrant geometry), and the stability limit of the suspended (top) state. These are evaluated using a macroscopic (sharp interface) model. Here, the energies of the collapsed, empty and suspended states are expressed respectively as
\begin{align}
	\widetilde{\Psi}_C &= \gamma_{sl}A_{sl}-\Delta P V,\label{psiC}\\ 
	\widetilde{\Psi}_E &= \gamma_{sv}A_{sv},\label{psiE} \\
	\widetilde{\Psi}_{S} &= \gamma_{sl}A_{sl}+\gamma_{sv}A_{sv}+\gamma_{lv}A_{lv}-\Delta P V. \label{psiS}
\end{align}
In 3D, $A_{sl}, A_{sv},$ and $A_{lv}$ are the solid-liquid, solid-vapour and liquid-vapour interfacial areas respectively, \textit{V} is the liquid volume. In 2D, each of these quantities become the two dimensional analogues; the $\gamma$ therefore represent energies per unit length with $\Delta P$ expressed as an energy per unit area. 

The boundaries where T and C have equal energy are shown in the phase diagrams in Fig. \ref{fig:2dpostmin} (c) (4-5 boundary) and (d) (5-6 boundary), and were constructed through interpolating between adjacent simulation points. Approximating the liquid-vapour interface as planar, $\widetilde{\Psi}_C$ and $\widetilde{\Psi}_{S}$ were equated to yield an analytic expression for the phase boundary. This analytic expression is observed to agree with the simulation boundary to within an accuracy of 0.1\%; insignificant compared to the uncertainty introduced through the interpolation. 

For the reentrant geometry, the boundary where B and E have equal energy, shown in Fig. \ref{fig:2dpostmin} (d) (3-4 boundary), was also constructed through interpolating adjacent simulation points. This boundary was modelled by equating $\widetilde{\Psi}_E$ and $\widetilde{\Psi}_{S}$, now however $\widetilde{\Psi}_{S}$ is expressed for the B state and we again approximate the liquid-vapour interface as planar. The model is in excellent agreement with the simulation boundary, the maximum difference between the two being a reduced pressure difference of 0.04. 

The stability limit of T, shown in  Fig. \ref{fig:2dpostmin} (c) (3-4 boundary) and (d) (4-5 boundary),  can be evaluated through consideration of the $\theta_o$-$\Delta P_r$ coordinate at which it becomes energetically favourable for the liquid-vapour interface to slide down the structure. This yields the critical pressure, $\Delta P_c$, for a post of square cross section \cite{Zheng2005}. The same interpretation applied to the reentrant geometry accurately yields the suspended (top)-suspended (bottom) phase boundary. For both phase diagrams, the theoretical critical pressure is in agreement with the interpolated simulation results, the discrepancy between the two increases to 3\% at the extremes of $\Delta P_r$ as the planar interface assumption becomes inaccurate. 

\begin{figure}[!b]
\centering
\begin{tabular}{p{4cm} p{4cm}}
a) & b) \\
\includegraphics[width=0.2\textwidth]{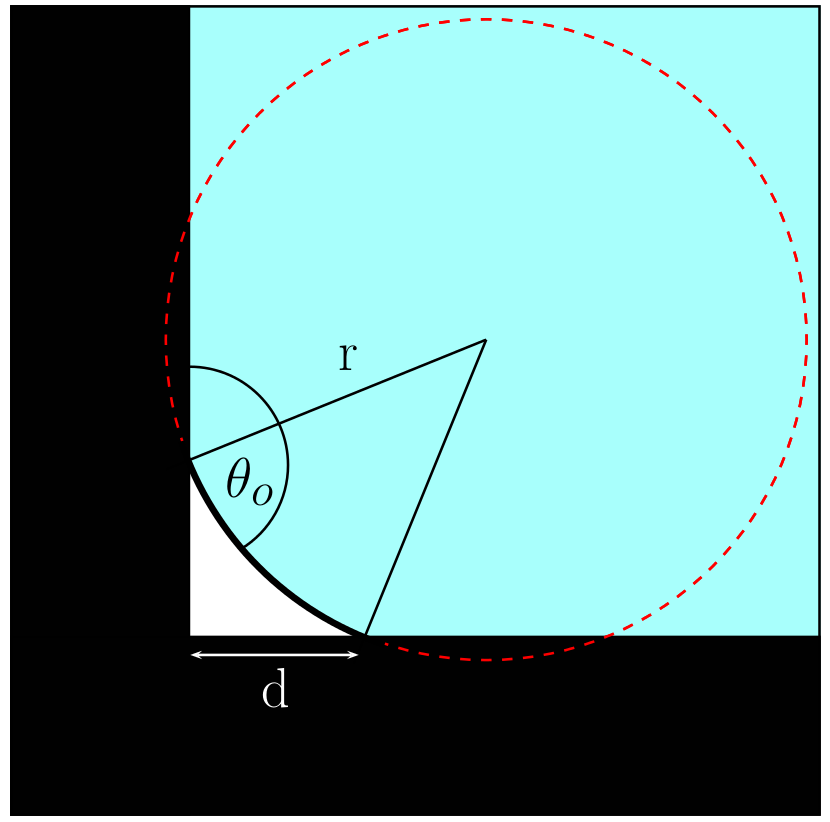} &
\includegraphics[width=0.2\textwidth]{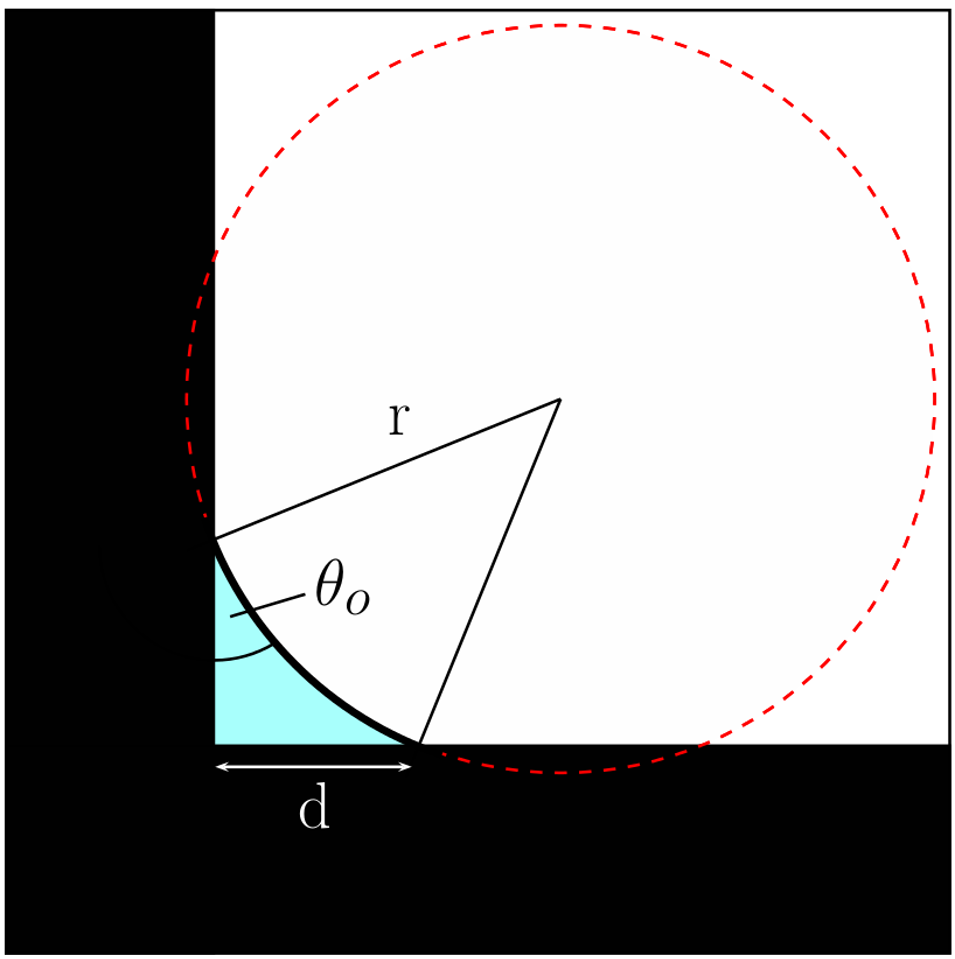}
\end{tabular}
\caption{(a) Geometric construction of the vapour cavity (white) at the base of the post (black), surrounded by the liquid phase (blue). The liquid-vapour interface is shown as a circular arc of radius \textit{r}, contacting the solid at the contact angle $\theta_o$, with lateral extend \textit{d}. (b) An equivalent model for the formation of a liquid condensate surrounded by the vapour phase.}
\label{fig:cavitation}
\end{figure}

Outside of the approximate range $45^o < \theta_o < 135^o$, cavitation and condensation dominate the phase behaviour, determining the stability limits for each of the minima shown in Fig. \ref{fig:2dpostmin} (a) and (b). As a first-order approximation, a sharp interface model is again assumed for the formation of vapour cavities and liquid condensates at the concave corners of the solid surfaces, illustrated in Fig. \ref{fig:cavitation} (a) and (b) respectively.

For the formation of a vapour cavity at the base of a post, the model assumes the detached liquid vapour interface to have constant curvature, described by a circular arc of radius \textit{r}. The condition is imposed that the liquid-vapour interface must contact the solid surface at the intrinsic contact angle $\theta_o$. The change in energy upon forming the vapour cavity (relative to the fully collapsed state) can therefore be expressed in terms of $\theta_o$ and \textit{d}, the maximum lateral extend of the cavity,
\begin{equation}
\Delta \widetilde{\Psi} = \left[ \frac{\sqrt{2}\gamma_{lv}d}{\sin\alpha}+\frac{\Delta P d^2}{2\sin^2\alpha} \right] \left[\alpha+\sin\alpha \left(\sin \alpha - \cos \alpha \right) \right], \label{eqn:vapourcav}
\end{equation}    
where $\alpha = \frac{3\pi}{4}-\theta_o$. Choosing to express the energy change in terms of $\alpha$ means that as $\theta_o \rightarrow \frac{3\pi}{4}, \alpha \rightarrow 0$. Small angle theorems may therefore be used to recover the energy change for a planar interface (when $r \rightarrow \infty$ but \textit{d} remains finite).

For $\alpha < 0, \Delta P < 0$, it is observed that $\Delta \widetilde{\Psi} < 0$, leading to a limitless increase in \textit{d}, and hence the instability of the collapsed state. For $\alpha > 0, \Delta P > 0$, the formation of a vapour cavity is energetically unfavourable. However, for $\alpha > 0, \Delta P < 0$, an energy barrier exists to unbounded cavitation such that both the empty and collapsed states are able to coexist. For $\alpha < 0, \Delta P > 0$ or  a vapour cavity of bounded extent is formed, with 
\begin{equation}
d=-\frac{\gamma_{lv}\sqrt{2}\sin \alpha}{\Delta P}. \label{eqn:dcon}
\end{equation}
Therefore, it is assumed that when \textit{d} is greater than half the distance between adjacent posts, the vapour cavities between proximal posts merge to lift the interface in a barrierless process. This condition forms the model collapsed stability limits shown in Figs. \ref{fig:2dpostmin} (c) and (d), and is accurate for the approximate range $-1.5 < \Delta P_r < 1.5$. For $\Delta P_r < -1.5$, cavitation is expected to occur on the flat surfaces, making the empty state monostable (regions 8 and 9 in Figs. \ref{fig:2dpostmin} (c) and (d) respectively). For $\Delta P_r > 1.5$, deviations from the sharp interface model presented are observed as an extension of the range of stability of regions 6 and 7 in Figs. \ref{fig:2dpostmin} (c) and (d) to lower pressures. Within these regions, the shape of the vapour cavity is significantly influenced by the interactions at the three-phase contact line \cite{Starov2007}, imaged in the Supplementary Information. Thus, a negative line tension operates to reduce the cavity size \textit{d} from that predicted by the sharp interface model in Eq. \eqref{eqn:dcon}, which neglects these contributions. 

At low $\theta_o$, the stability limit of the suspended (bottom) and empty states are derived in a similar manner to the collapsed stability limit, and is illustrated by the geometric construction in  Fig. \ref{fig:cavitation} (b). In this case, a liquid condensate is modelled at the base of the post, surrounded by the vapour phase. The energy change for this condensation is therefore expressed as
\begin{equation}
\Delta \widetilde{\Psi} = \left[ \frac{\sqrt{2}\gamma_{lv}d}{\sin\delta}+\frac{\Delta P d^2}{2\sin^2\delta} \right] \left[\delta-\sin\delta \left(\cos \delta + \sin \delta \right) \right], \label{eqn:liquidcon}
\end{equation} 
where $\delta = \frac{\pi}{4}-\theta_o$. When $\delta > 0, \Delta P > 0$, it is energetically favourable for a condensate to form which fills the simulation volume. For $\delta < 0, \Delta P <0$, condensation does not occur. When $\delta < 0, \Delta P > 0$, an energy barrier exists to nucleation such that the collapsed and empty states are able to coexist, whereas for $\delta > 0, \Delta P < 0$, a condensate of finite \textit{d} is formed, where
\begin{equation}
d=-\frac{\gamma_{lv}\sqrt{2}\sin \delta}{\Delta P}. \label{eqn:dliq}
\end{equation}
Once again, when \textit{d} is greater than half the post separation, it is assumed that the condensates between adjacent posts merge to fill the simulation volume with liquid in a barrierless process. This condition forms the model empty stability limits shown in Figs. \ref{fig:2dpostmin} (c) and (d). Because an identical condensate merger process is expected to take place in the suspended (bottom) state, the empty stability limit is also the stability limit of the suspended (bottom) state shown in Fig. \ref{fig:2dpostmin} (d). In an identical manner to the increased stability range of states V and D, states N and $\rm{D}_i$ in region 2 of Figs. \ref{fig:2dpostmin} (c) and (d) experience a stability range increase to lower pressure magnitudes to that expected from the sharp interface model presented. In the low-$\theta_o$ regime, line tension effects again become non-negligible, with a positive line tension reducing the condensate size \textit{d} relative to Eq. \eqref{eqn:dliq}. 

Comparing the two phase diagrams in Figs. \ref{fig:2dpostmin} (c) and (d), we conclude that in two dimensions, the dominant effect of the reentrant structure is to maintain a suspended state over a greater range in the phase diagram than the 2D post allows. However, the dimensionality of the system and $90^\circ$ angles of the solid corners mean that both the post and reentrant structures experience similar empty and collapsed stability limits with respect to cavitation and condensation. Despite the ability of the 2D reentrant structure to support a suspended (bottom) state, identical condensation effects to those prevalent in the empty state at low contact angles mean that it is only possible to suspend liquids with $\theta_o$ significantly less than $45^\circ$ at large negative pressures.

\begin{figure*}[!ht]
\centering
\includegraphics[width=\textwidth]{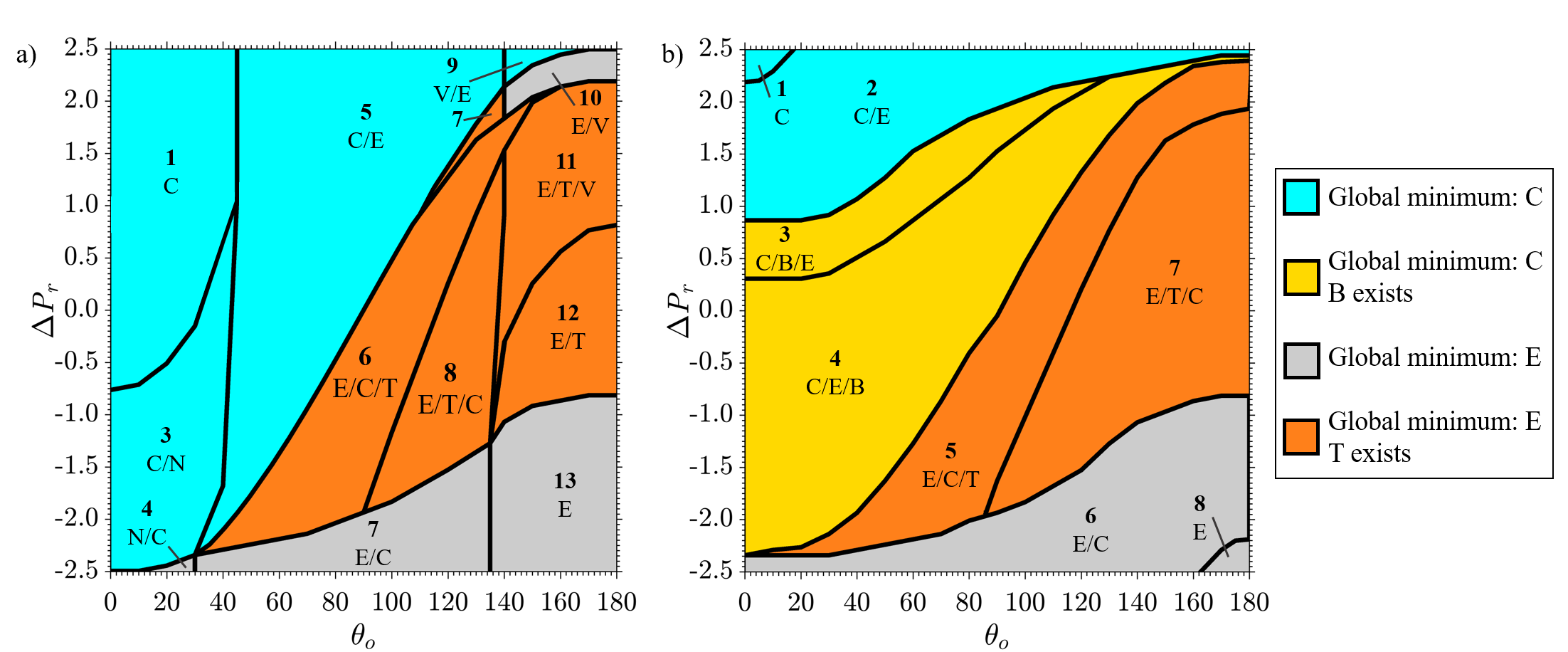}
\caption{Phase diagrams for the 3D post (a) and 3D reentrant geometries (b). Numbered regions indicate the stable states, listed in order of increasing energy }
\label{fig:3dphasediagrams}
\end{figure*}

In three dimensions, the free energy minima of the posts and reentrant structures are also accurately represented by the phase field images of the 2D structures in Fig. \ref{fig:2dpostmin} (a) and (b). In 3D however, these images represent cross sections through the system. The phase diagrams are shown in Fig. \ref{fig:3dphasediagrams} (a) and (b) respectively.

The 3D post phase diagram is qualitatively similar to the 2D equivalent, whereas significant differences exist between the 2D and 3D reentrant phase diagrams. The differences between the phase diagrams are dominated by the cavitation and condensation effects, in particular the change in liquid-vapour interface morphology concomitant with the change from 2D to 3D. 
 
Around the perimeter of the 3D post and reentrant pillar, the liquid-vapour interface of the cavity/condensate forms a surface of constant mean curvature, which can be expected to be similar to an unduloid distorted by the square post geometry. The first observed effect of this is to reduce the pressure range over which the suspended states are stable in 3D, relative to 2D. Secondly, whereas in 2D the cavity extent \textit{d} is independent of the pillar width, in 3D \textit{d} is proportional to the pillar width. Crucially for the reentrant geometry used here, the pillar is sufficiently narrow that cavity and condensate formation is not observed across all $\theta_o$ at the level of the simulation resolution. Therefore, an effective method to stabilise a suspended state with respect to condensation is to fabricate a reentrant surface texture in which the pillar width is minimised. An increase in the suspended-collapsed coexistence range has been observed previously using a diffuse interface model, but for a droplet on a square array of reentrant posts \cite{Zhang2014}. However, the suspended and collapsed states were not stable over all $\theta_o$. In this system, the ratio between the reentrant pillar width and the post width was only 0.5, compared to 0.17 here, meaning that the maximisation of the coexistence range was not effectively realised.

A consequence of reducing the occurrence of cavitation in the 3D reentrant geometries is that the suspended states always coexist with the collapsed state. This is in contrast to region 12 of Fig. \ref{fig:3dphasediagrams} for the post structure where, neglecting the empty state, the suspended state is monostable, even at positive pressures. The occurrence of a monostable suspended state on a square array of posts has been recently shown to occur experimentally for hierarchically textured, superhydrophobic surfaces \cite{Li2016}. The addition of a nanoscale texture to the microscale structuring effectively accesses contact angles in excess of $135^\circ$ (for surfaces without the microscale texture). Here we have shown that such surfaces represent the perfect robustness of the suspended state. One caveat however is that in real systems prepared in the collapsed state, the recovery of the suspended state through growth of a vapour cavity will be significantly hindered by the diffusion rate of gases through the liquid phase, unless other mechanisms to increase the gas volume are employed \cite{Lee2011}.

\begin{figure}[!ht]
\centering
\includegraphics[width=0.45\textwidth]{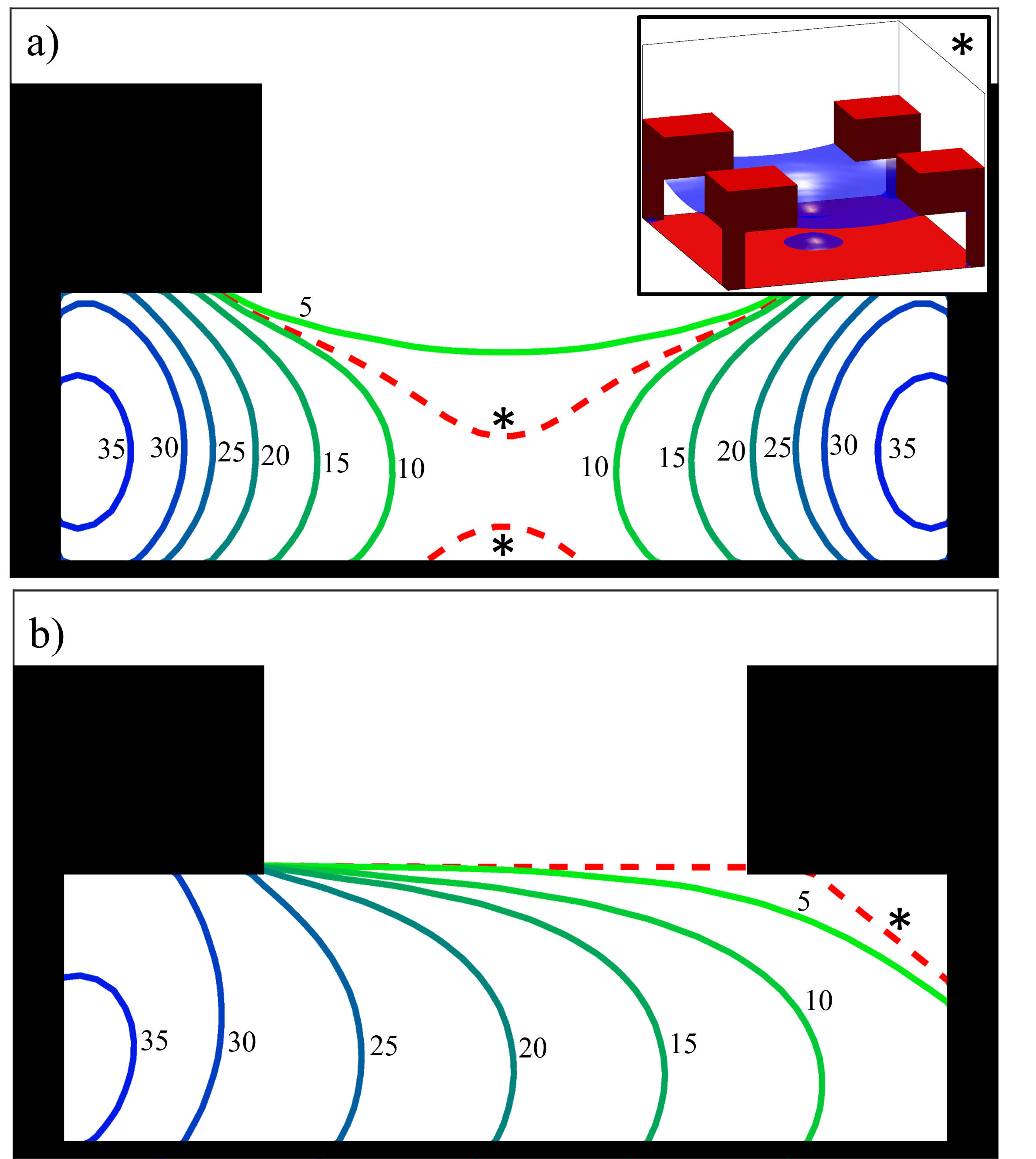}

\caption{(a) MEP for the collapse transition of the 2D reentrant structure at $\theta_o = 50^\circ, \Delta P_r = 0.15$, see video S1. (b) Diagonal cross section of the MEP for the collapse transition of the 3D reentrant structure at $\theta_o = 20^\circ, \Delta P_r = 0.15$, see video S2. The liquid-vapour ($\phi = 0$) interface is shown at 5-image intervals. The interface at the transition state (red, dashed) is also labelled ($\ast$).}
\label{fig:MEPcon}
\end{figure}

A second consequence of the increased suspended-collapsed coexistence range for both 2D and 3D reentrant geometries, relative to posts, is that at low contact angles condensates are observed to participate in the collapse transition pathways. The MEP for such a condensate-incorporating mechanism for the 2D reentrant structure is shown in Fig. \ref{fig:MEPcon} (a). Here, the transition is initiated through condensation of the liquid under the cap structure with an early transition state occurring when the growing condensate reaches the cap edge. At this point, the liquid-vapour interfacial area is maximal.

In the 3D reentrant structures, a condensate also critically affects the transition pathway, shown in Fig. \ref{fig:MEPcon} (b). Unlike in the 2D case, here the liquid-vapour interface first sags underneath the cap, remaining pinned only on the centre of each cap edge. At the same time, liquid condenses on the base of the system at the midpoint between diagonally adjacent pillars. At the transition state this sagging interface and the condensate coalesce. The interface then depins from cap edges and slides under the cap. In the final stages of the collapse mechanism, the vapour forms a toroidal ring around the reentrant pillar. 

Overall, it is shown that in both 2D and 3D systems at low contact angles, condensates not only participates within the MEP, but crucially affects the liquid-vapour interface configuration at the transition state. Therefore, in designing reentrant surfaces which exhibit a robust suspended state, condensation effects cannot be neglected for highly wetting liquids.

\section{Transition mechansisms in three dimensions} \label{sec:reen}
Even when condensation or cavitation effects are not dominant, the shape of the reentrant geometry is still observed to critically influence the collapse transition pathway. Here, we show the existence of two fundamental transition mechanisms of reentrant structures, and compare the properties of the MEPs of each as the contact angle is varied.

\begin{figure*}[!ht]
\centering
\includegraphics[width=\textwidth]{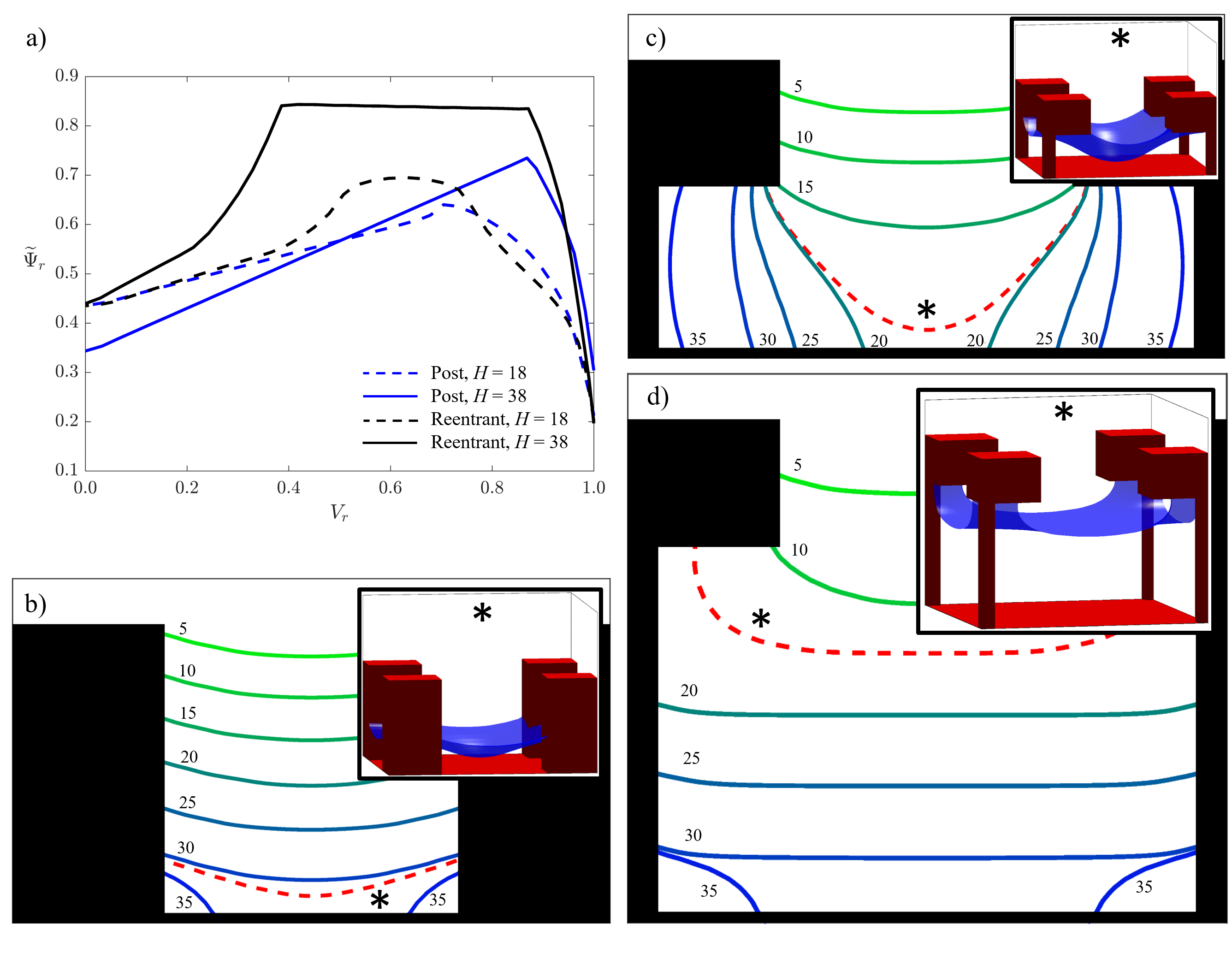}
\caption{(a) Minimum energy pathways between suspended and collapsed states of the four sample structures (labelled); $\theta_o =110^\circ, \Delta P_r = 0.15$. The progress of the transition is monitored through evaluation of the reduced liquid volume, $V_r$. (b-d) Diagonal cross section of the MEP for the 3D collapse transitions at $\theta_o = 110^\circ, \Delta P_r = 0.15$ for the post (b), video S3, short reentrant (c), video S4, and tall reentrant geometries (d), video S5. The liquid-vapour ($\phi = 0$) interface is shown at five-image intervals. The interface at the transition state (red, dashed) is also labelled ($\ast$). Inset: 3D visualisations of the interface (blue) at the transition state.
}
\label{fig:3dmech}
\end{figure*}

The minimum energy pathways are compared for the 3D post and reentrant structures between the suspended and collapsed states. First, a comparison is made between the energetic profiles along the MEPs of the two geometries at both tall and short heights, shown in Fig. \ref{fig:3dmech} (a). Here, each transition takes place at $\theta_o = 110^\circ, \Delta P_r = 0.15$. We chose to simulate transitions at $\Delta P_r = 0.15$ throughout this work, in order to correspond to experimental scenarios in which a textured surface is submerged to a depth of approximately 2cm under water, when a length scale \textit{G} = 1 $\mu \rm{m}$ is chosen. The transitions are followed through consideration of the reduced liquid volume, $V_r$,
\begin{equation}
V_r=\frac{V-V_{S}}{V_C-V_S}.
\end{equation}
Here, \textit{V} is the absolute liquid volume within the system at a point along the MEP. $V_C$ and $V_S$ are the liquid volumes in the collapsed and suspended states respectively. 

Let us focus on the 3D post geometries. In agreement with previous studies, the wetting transition for both the short and tall posts is observed to take place via the depinning mechanism, in which the interface depins from the top of the post before sliding to the base of the system \cite{Butt2014,Patankar2010,Jung2007,Kusumaatmaja2008, Tuteja2008,Pashos2015,Shahraz2014,Bormashenko2015}. This is illustrated in Fig. \ref{fig:3dmech} (b). The transition state occurs at the point where the liquid-vapour interface first contacts the base of the system, followed by rapid completion of the collapse. This behaviour is in excellent agreement with theoretical predictions \cite{Shahraz2014,Pashos2016,Zhang2014,Ren2014,Cai2016} and experimental observations \cite{Reyssat2008,McHale2005}. Throughout the range of pressures and contact angles, the alternative sagging mechanism is not observed due to the large post height to separation ratio (even for what are termed short posts considered here).  

For the  3D reentrant geometries, the striking observation is made that upon increasing the pillar height, there is a significant change in transition mechanism. This is evidenced in the change in shape of the MEP energetic profile in Fig. \ref{fig:3dmech} (a) upon increasing \textit{H} from 18 to 38. The transition pathways for each height are visualised in Fig. \ref{fig:3dmech} (c) and (d) respectively.

For the short reentrant geometry, shown in Fig. \ref{fig:3dmech} (c), for all $\theta_o$, the liquid-vapour interface remains pinned at the centre of the bottom edge of the cap structure whilst the interface sags in the diagonal separation between the structures. As the interface approaches the base of the system, it becomes pinched in the centre to achieve the transition state. Following this, the interface depins completely from the cap edge, before sliding underneath to complete the transition.

For the tall structures however, a significantly different mechanism is observed, shown in Fig. \ref{fig:3dmech} (d). As with the short structure, the first stage of the transition is completed by the liquid-vapour interface sagging below the cap, whilst remaining pinned on the centre of the cap edges. In the tall geometry however, eventually the sagging process ceases completely, and the impinging liquid extends laterally under the cap. Only after contacting the pillar does the interface continue to slide down towards the system base. In this sliding region, the shape of the interface remains constant, leading to the plateau in the MEP energetic profile seen in Fig. \ref{fig:3dmech} (a). Thus, it is not true that the transition is rapidly completed after the interface first contacts the solid surface beneath the top of the structure, in direct contrast to the post transition mechanisms.

We therefore distinguish two distinct collapse mechanisms on reentrant geometries: Base Contact, Fig. \ref{fig:3dmech} (c), and Pillar Contact, Fig. \ref{fig:3dmech} (d). The key characteristics of both mechanisms have been described using the short and tall reentrant posts respectively. For intermediate post heights, both base contact and pillar contact modes are MEPs, the post height and contact angle determining which mode has the lowest transition state energy.

It is predicted that the advent of transparent reentrant surface textures \cite{Lee2013} may enable these mechanisms to be distinguished experimentally via similar optical techniques used to elucidate the post transition mechanism \cite{Lv2014, Lv2015, Rathgen2010}.

\begin{figure*}[!ht]
\centering
\begin{tabular}{p{0.5\textwidth} p{0.5\textwidth}}
a) & b) \\
\includegraphics[width=0.45\textwidth]{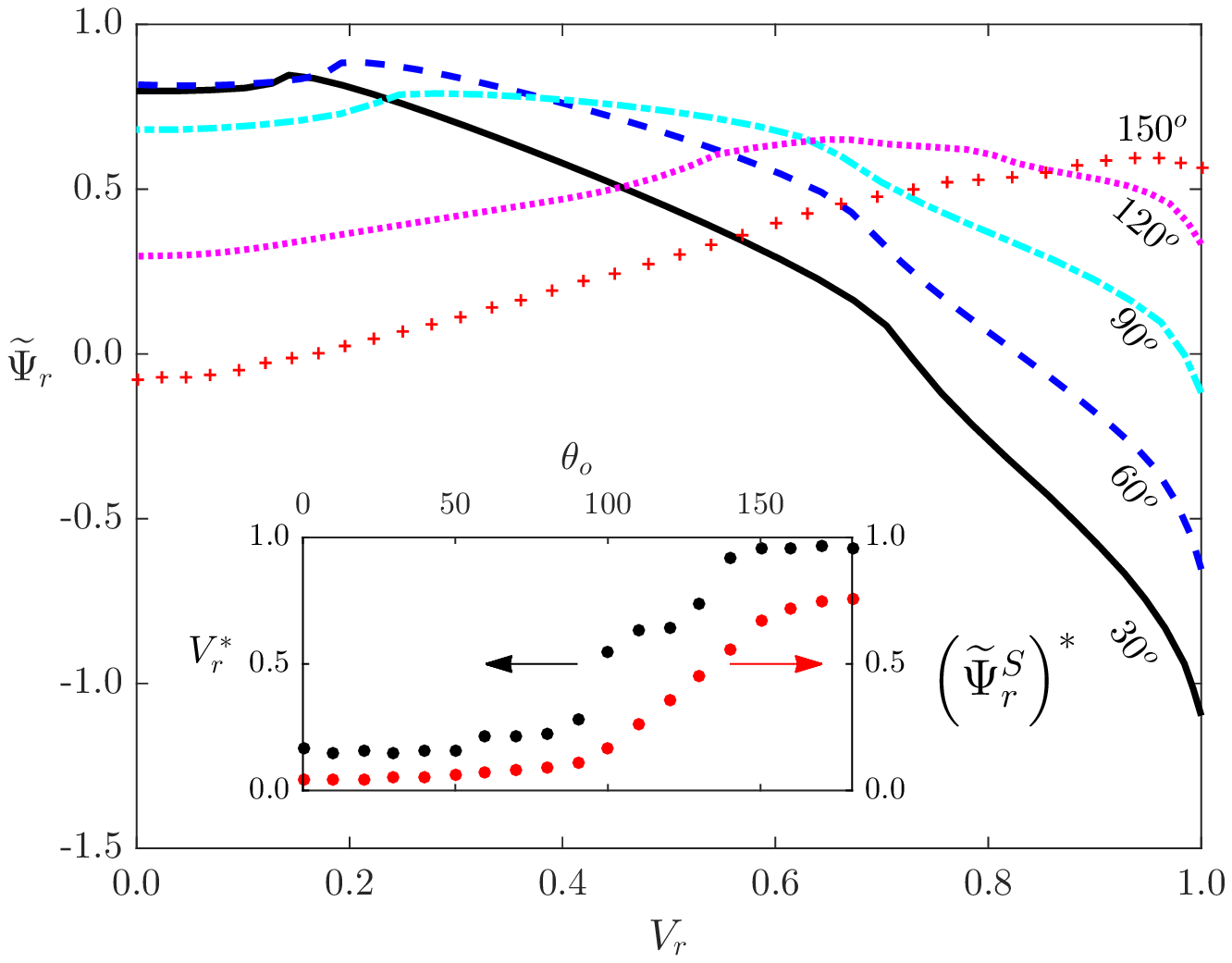} &
\includegraphics[width=0.45\textwidth]{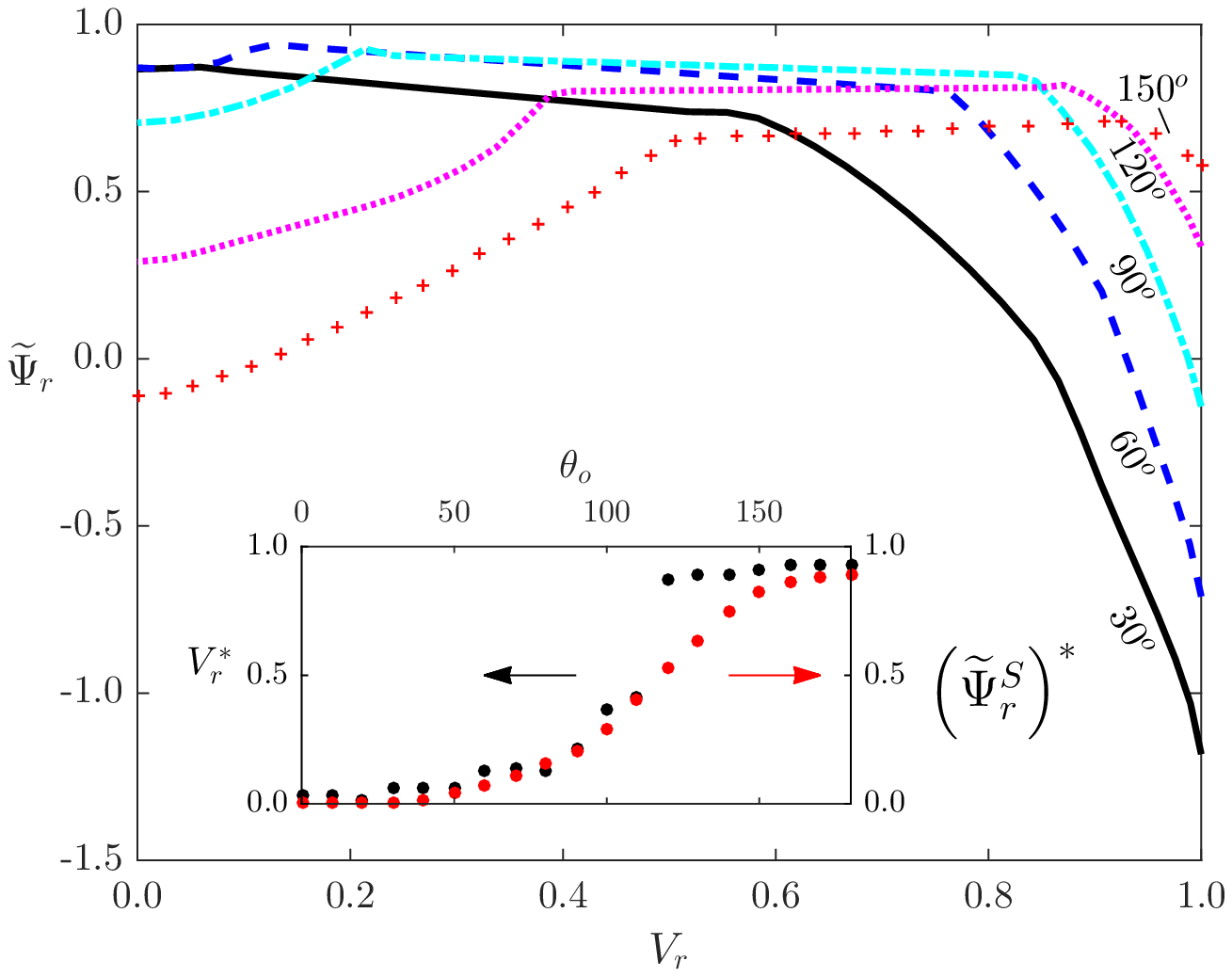} 
\end{tabular}
\caption{The influence of $\theta_o$ (labelled) on the energetic profiles of the base contact (a) and pillar contact (b) mechanisms. All transitions are carried out at $\Delta P_r = 0.15$. Inset: the variation of the position (black) and energy (red) of the transition state (relative to the suspended minimum free energy) as $\theta_o$ is varied.}
\label{fig:3dMEPCA}
\end{figure*}

The variation of the energetic profiles of the MEPs as $\theta_o$ is changed is compared for both mechanisms at $\Delta P_r =0.15$. Fig. \ref{fig:3dMEPCA} (a) shows the behaviour for the short reentrant geometries (Base Contact), whilst Fig. \ref{fig:3dMEPCA} (b) shows the behaviour for the tall geometries (Pillar Contact). Condensates are observed to participate in the transitions for $\theta_o < 50^\circ$.

For the base contact mode, as $\theta_o$ is increased, the energy of the transition state relative to the suspended state $\left( \Delta \widetilde{\Psi}_r^S \right)^*$ evidences a large increase between $90^\circ$ and $140^\circ$. At low contact angles, the transition state occurs when the sagging liquid-vapour interface first contacts the system base. However, for $\theta_o \geq 140^\circ$, although the same base contact mechanism is followed, the transition state occurs immediately prior to the interface contacting the pillar at the end of the transition. A crossover regime is observed close to $\theta_o = 130^\circ$, in which the transition state occurs between the interface contacting the base and the pillar. This change in transition state position are evidenced by and increase in the reduced liquid volume at the transition state, $V_r^*$.

The tall reentrant geometry also evidences a jump in both the transition state energy and location between $\theta = 110^\circ$ and $\theta = 120^\circ$. In these systems, as the interface slides down the pillar in the plateau region of the MEP, for $\theta \geq 120^\circ$ the energy decrease for increasing the volume of liquid in the system is exceeded by the energy 
penalty of increasing the liquid-solid contact area. Thus, for  $\theta < 120^\circ$, the transition state occurs immediately prior to the interface first contacting the pillar. For $\theta \geq 120^\circ$ however, the transition state is immediately prior to to interface contacting the base of the system. 

To complete the reentrant transition mechanism discussion, the effect of $\Delta P_r$ on the MEP energetic profile was investigated, and is shown in the Supplementary Information. Unlike when varying $\theta_o$, no significant change in the shape of the MEP is observed upon changing $\Delta P_r$. The only effect is to change the free energy of each image along the MEP in proportion to the volume of liquid present. This behaviour is observed for all systems (2D, 3D, post and reentrant) in the pressure range simulated.

\section{Transition symmetry} \label{sec:sym}
It has been observed previously, using molecular dynamics and sharp interface models, that in two dimensions, the reflection symmetry of the system is broken during the collapse transition \cite{Giacomello2012, Giacomello2015, Amabili2015,Amabili2016}. In this section we first corroborate these findings, and link the disparate length scales of the molecular dynamics and sharp interface models with our diffuse interface model. The geometric influence of when and how the symmetry is broken along the MEP is then elucidated. Finally, in contrast to the two collapse mechanisms observed in 3D reentrant geometries, we observed three distinct mechanisms in 2D.

\begin{figure*}[!ht]
\centering
\includegraphics[width=\textwidth]{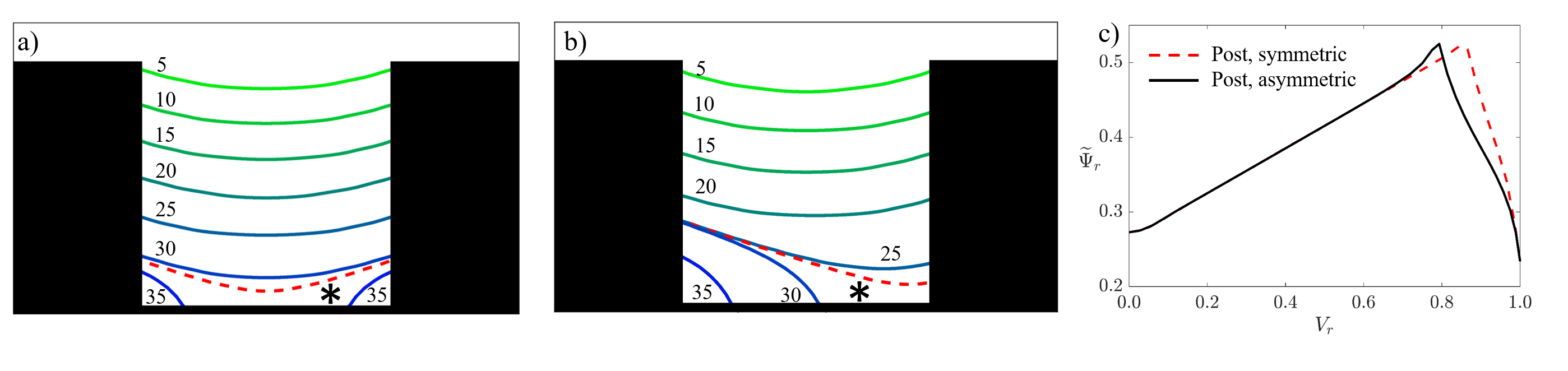}
\caption{(a, b) Liquid-vapour interfacial profile ($\phi = 0$) for the 2D post collapse transition shown every five images with the transition state highlighted (red, dashed). $\theta_o = 100^\circ, \Delta P_r = 0.15$. (a) Symmetric mechanism. (b) Asymmetric mechanism, video S6. (c) MEP energetic profiles for the symmetric (red, dashed) and asymmetric (black) collapse mechanisms.}
\label{fig:2dMEPsym}
\end{figure*}

Beginning the discussion with 2D square posts (or equivalently square wells), molecular dynamics and sharp interface models have previously shown that an asymmetric transition pathway is observed for the collapse mechanism \cite{Giacomello2012, Giacomello2015}. Here however, two mechanisms are observed: a symmetric transition and and asymmetric transition, shown in Fig. \ref{fig:2dMEPsym} (a) and (b) respectively.

The symmetric path is obtained for symmetric initial DNEB strings, or strings with moderate asymmetry introduced. It is found that the asymmetric transition path can only be accessed via introducing an extreme asymmetry into the initial DNEB string. It is noted here that initialising the 3D systems with extremely asymmetric strings does not result in an asymmetric pathway. Experimental testing of the wetting transition found that an asymmetric transition pathway was never observed in a 3D circular well (unless impurities were present) \cite{Lv2015}. To explain the discrepancy between this experimental observation and the previous 2D simulations, it was suggested that although the asymmetric path may exist, during the experiment the motion of the liquid-vapour interface is along the symmetric reaction coordinate. Between equivalent points on the symmetric and asymmetric MEPs, it was further suggested that a high energetic barrier exists. From our 2D simulations, the stability of the symmetric MEP with respect to asymmetric perturbations corroborates this suggestion. However, in order to show this conclusively, it would be necessary to determine whether an asymmetric path is available within a 3D cylindrical well.

The energetic profiles for the symmetric and asymmetric post transition mechanisms are shown in Fig. \ref{fig:2dMEPsym} (c). For the majority of the collapse transition, both mechanisms share identical paths, particularly when the liquid-vapour interface initially slides down the post. In the vicinity of the transition state, the asymmetric mechanism evidences an earlier transition state of lower energy, compared to the symmetric path. However, the energy difference between the transition states of the two mechanisms is only approximately 0.5\%.

\begin{figure*}[!ht]
\centering
\includegraphics[width=\textwidth]{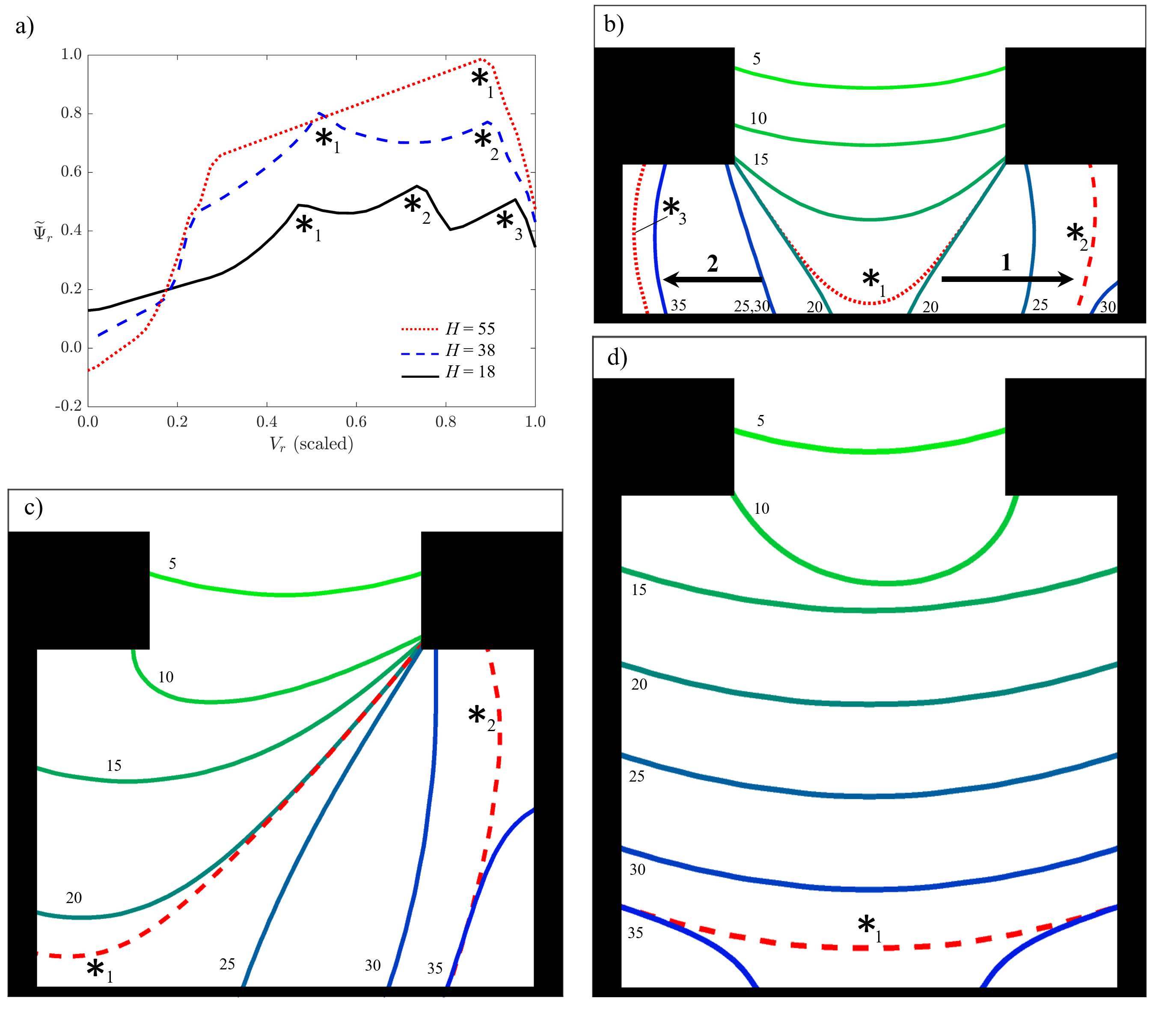}
\caption{(a) Energetic profiles along the MEPs of the 2D reentrant geometries at three pillar heights (labelled). (b-d) The liquid-vapour interface ($\phi=0$) at five-image intervals along the MEP, showing the local energetic maxima (red, dashed). $\theta=110^\circ, \Delta P_r = 0.15$. (b) Base Contact: \textit{H}=18, labelled arrows indicate the order in which each side of the system wets the solid structure, see video S7. (c) Hybrid: \textit{H}=38, video S8. (d) Pillar Contact: \textit{H}=55, video S9.}
\label{fig:2dmech}
\end{figure*}

For the 2D reentrant systems, an asymmetric collapse transition is obtained regardless of the initial DNEB string. As with the 3D reentrant geometries, the collapse mechanism is influenced by the pillar height, but in two dimensions, three collapse mechanisms are observed with differing numbers of local maxima along the MEP energetic profile, shown in Fig. \ref{fig:2dmech} (a). 

At \textit{H}=18, a base contact mechanism is observed, shown in Fig. \ref{fig:2dmech} (b). Initially, the interface sags symmetrically between the structures to contact the base of the system, defining the position of the first of three local maximum along the MEP, shown in Fig. \ref{fig:2dmech} (a). After this point, one side of the liquid-vapour interface remains pinned to the cap edge, whilst the other slides under the cap. Immediately prior to contacting the pillar, a second local energetic maximum (the transition state in this system) occurs. Finally, the pinned interface detaches, and also slides under the cap. Again, the point at which the interface contacts the pillar defines the third local energetic maximum.

At \textit{H}=38, unlike with the 3D case, the pillar contact mechanism is not observed. Instead a hybrid mechanism is evidenced in Fig. \ref{fig:2dmech} (c), showing aspects of both Base Contact and Pillar Contact. Initially, one side of the liquid-vapour interface slides under the cap by lateral deformation to contact the pillar, whilst the other remains pinned at the cap edge. Unlike in the 3D case, this does not represent a transition state, but does evidence a kink structure in the energetic profile in Fig. \ref{fig:2dmech} (a). As the interface slides down the pillar on one side only, the reflection symmetry of the system is increasingly violated. The first of two local maxima occurs when the interface contacts the base of the system (the transition state for this system). Only after this has occurred can the interface depin from the cap edge and slide under the cap. The second local maximum is observed immediately prior to the interface contacting the pillar.  

Only by increasing the post height to \textit{H}=55 can a pillar contact mechanism be accessed, shown in Fig. \ref{fig:2dmech} (d). It is noted that in order to simulate this pillar height, $N_z = 80$ was used. Here, the interface sags under the cap, to contact the pillar asymmetrically. As with the hybrid mechanism, this does not represent the transition state, but a kink in the MEP energetic profile. Following this, the interface slides symmetrically down the pillar to contact the system at the base (defining the transition state). The energetic profile for this process in Fig. \ref{fig:2dmech} (a) evidences a plateau-like structure seen in the 3D MEPs. The gradient in 2D is exaggerated relative to the 3D case however, due to the relative increase in energetic contribution from the interfacial energies relative to the $\Delta P-V$ term. Unlike in the previous two mechanisms, the reflection symmetry in 2D Pillar Contact is never strongly broken. It is further noted that at intermediate pillar heights between the three cases presented, two transition pathways are accessible, with the MEP of lowest transition state determined by the pillar height.

The existence of three collapse mechanisms with very different energetic MEP profiles leads to the general conclusion that 2D systems are more strongly influenced by surface geometry than equivalent 3D systems. This is expected, due to the relative increase in $\frac{A_{sl}}{V}$ in 2D systems relative to the 3D equivalents.

\section{Conclusions}
Overall, we have developed a phase field model designed to selectively probe the influence of the system surface geometry on the minimum free energies, and collapse transition pathways. This has been achieved for 2D and 3D systems, featuring square posts and reentrant structures. Within this description, three principal outcomes which significantly influence future surface design are highlighted.

Firstly, condensation and cavitation processes were shown to be able to critically influence both the minimum stability ranges in $\theta_o$ and $\Delta P_r$, and the collapse mechanisms. A model was developed which accurately predicted the stability ranges of the 2D and 3D posts, when the liquid-vapour interface had a constant profile around the perimeter of the surface structure. For the 3D reentrant posts, this approximation was not correct; the corners of the square pillar enabled the empty, collapsed and suspended states to coexist across the entire phase diagram. This was due to the formation of condensates and cavities around the base of the pillar of significantly different shape, and much larger liquid-vapour interface area than assumed in the model.  

The second principal outcome was the description of two dominant collapse mechanisms for the reentrant geometries: 1) the base-contact mode, characterised by a sagging interface  first contacting the base of the system, 2) the pillar contact mode, characterised by the impinging liquid-vapour interface deforming laterally under the cap to contact the pillar first. In this work, each mechanism could be sampled by changing the pillar height. 

The final principal outcome was the observation that 2D systems undergo significantly different collapse mechanisms than their three dimensional counterparts, particularly regarding the prevalence of asymmetric pathways. Three collapse mechanisms were characterised for the 2D reentrant geometries: the 2D asymmetric equivalents of the base contact and pillar contact modes, and a third highly asymmetric mechanism which is a hybrid of the previous two. Furthermore, the pillar heights at which each mechanism is operative differs from the 3D equivalents.

Overall, it is anticipated that the fundamental geometrical effects presented in this work will enable the rational and targeted design of robust superamphiphobic surfaces.

\section{Acknowledgements}

We thank Dr. M. Wagner, Dr. Y. Gizaw, Dr. C. Semprebon and M. S. Sadullah for fruitful discussions. We acknowledge Procter and Gamble (P$\&$G) for funding.

\bibliography{References}
\end{document}

\maketitle

\section{Total free energy: effect of the pressure term}
In total, the free energy of the system is described by
\begin{equation}
\Psi-\Psi_o = \int_V \left(\psi_b + \frac{\epsilon}{2} \vert \nabla \phi \vert^2 \right)dV + \int_S \psi_s dS - \Delta P \int_V \frac{\phi+1}{2} dV, \label{eqn:psi}
\end{equation}
where $\psi_b = \frac{1}{\epsilon}\left(\frac{1}{4}\phi - \frac{1}{2}\phi \right)$.

The effect of the pressure term is to shift the bulk value of $\phi$ relative to the $\Delta P =0$ scenario, determined by minimising $\Psi-\Psi_o$ in the absence of any interfaces. Thus, for an isotropic, homogeneous system,
\begin{equation}
\phi_o^3-\phi_o-\frac{\Delta P \epsilon}{2} = 0.
\end{equation}
This admits three complex solutions for:
\begin{equation}
\phi_o=
\begin{cases}
\frac{1}{3Q}+Q \quad \left(\rm{= +1 \quad when}\quad \Delta \textit{P} = 0\right),\\
-\frac{1}{6Q}-\frac{Q}{2}+\frac{\sqrt{3}}{2}\left(\frac{1}{3Q}-Q\right)i \quad \left(\rm{= -1 \quad when}\quad \Delta \textit{P} = 0\right), \\
-\frac{1}{6Q}-\frac{Q}{2}-\frac{\sqrt{3}}{2}\left(\frac{1}{3Q}-Q\right)i \quad \left(\rm{= 0 \quad when}\quad \Delta \textit{P} = 0\right),
\end{cases}
\end{equation}
where
\begin{equation}
Q=\left[\frac{\Delta P \epsilon}{4}+\sqrt{\left(\frac{\Delta P \epsilon}{4}\right)^2-\frac{1}{27}}\right]^{\dfrac{1}{3}}.
\end{equation}
Only the first and second solution yield free energy minima, the third yields the system maximum. Since \textit{Q} is itself complex, two real solutions for $\phi_o$ exist only  in the range $\vert \phi_o \vert > \frac{\sqrt{3}}{3}$, corresponding to the approximate pressure range $\vert \Delta P_r \vert < 49.1$, outside of this range only a single real minimum is admitted.

\section{System discretisation}

To calculate approximate numerical solutions to $\Psi-\Psi_o$, Eq.\eqref{eqn:psi} is discretized term by term as follows:
\begin{align}
\int_V\psi_bdV&=\sum_{ijk}\frac{1}{\epsilon}\left(\frac{1}{4}\phi_{ijk}^4-\frac{1}{2}\phi_{ijk}^2\right)G^3, \\
\int_V \frac{\epsilon}{2} \vert \nabla \phi \vert^2 dV&=\sum_{ijk}\frac{\epsilon}{2}\left(\left(\frac{\partial \phi_{ijk}}{\partial x}\right)^2 + \left(\frac{\partial \phi_{ijk}}{\partial y}\right)^2 + \left(\frac{\partial \phi_{ijk}}{\partial z}\right)^2\right)G^3, \label{eqn:nabla} \\ 
\int_S \psi_S dS &= \sum_{\rm{surface}}-h\phi_{ijk}G^2, \\ 
\Delta PV_l &= \Delta P \sum_{ijk} \frac{\phi+1}{2}G^3.
\end{align}
Here, the spatial derivatives are approximated to second order accuracy and are calculated as, for example,
\begin{equation}
\left(\frac{\partial \phi_{ijk}}{\partial x}\right)^2=\frac{1}{2G^2}\left(\left(\phi_{(i+1)jk}-\phi_{ijk}\right)^2+(\left(\phi_{(i-1)jk}-\phi_{ijk}\right)^2\right).
\end{equation}
A similar treatment is employed in the \textit{y} and \textit{z} directions.

Periodic boundary conditions are enforced in the \textit{x} and \textit{y} directions, whereas the \textit{z}-gradient at $k=N_z$ is fixed as zero. This is equivalent to having a surface which does not interact with the fluid phases in any other way than to prevent them interacting with the underside of the solid structure of interest. The boundary conditions for the gradient at the solid surface are also imposed in terms of the directional derivative of $\phi$:
\begin{equation}
\widehat{\underline{\textbf{n}}}\cdot\nabla\phi_{ijk}=-\frac{h}{\epsilon}.
\end{equation} 
Here, the same discretization of $\nabla\phi_{ijk}$ is used as in Eq. \eqref{eqn:nabla}. Since these surface gradient terms contribute a constant to the free energy we are free to neglect its contribution in Eq. \eqref{eqn:nabla}. $\widehat{\underline{\textbf{n}}}$ is the unit vector normal to the surface element. At edges and vertices of the structure, the direction of the normal is defined as the average of the normals of the incident planes. This scheme is slightly different to others employed previously \cite{Dupuis2005}, in which the node states are chosen according to the permitted directions in which derivatives can be taken. 

Overall therefore, the gradient in the free energy with respect to the order parameter at each node can be computed:
\begin{equation}
\frac{d\widetilde{\Psi}}{d\phi_{ijk}}=\frac{d\Psi_i}{d\phi_{ijk}}+\frac{d\Psi_s}{d\phi_{ijk}}-\Delta P \frac{dV}{d\phi_{ijk}}, \label{eqn:psigradient}
\end{equation}
where
\begin{align}
\frac{d\Psi_i}{d\phi_{ijk}}&=\frac{G^3}{\epsilon}\left(\phi_{ijk}^3-\phi_{ijk}\right) \nonumber \\ 
&-\epsilon G\left(\left(\phi_{(i+1)jk}-\phi_{ijk}\right)+(\left(\phi_{(i-1)jk}-\phi_{ijk}\right)\right) \nonumber \\ 
&-\epsilon G\left(\left(\phi_{i(j+1)k}-\phi_{ijk}\right)+(\left(\phi_{i(j-1)k}-\phi_{ijk}\right)\right) \nonumber \\ \label{eqn:dpsidphi}
&-\epsilon G\left(\left(\phi_{ij(k+1)}-\phi_{ijk}\right)+(\left(\phi_{ij(k-1)}-\phi_{ijk}\right)\right), \\ 
\frac{d\Psi_s}{d\phi_{ijk}} &=-hG^2, \quad \rm{if \, \phi_{ijk} \, is \, a \, boundary \, node} ,\\
\frac{dV}{d\phi_{ijk}} &= \frac{G^3}{2}. 
\end{align}
Note that for surface nodes, a spatial derivative component of Eq. \eqref{eqn:dpsidphi} is zero if the direction in which the derivative is taken is parallel to  $\widehat{\underline{\textbf{n}}}$.

\begin{figure}[!b]
\centering
\includegraphics[width=0.6\textwidth]{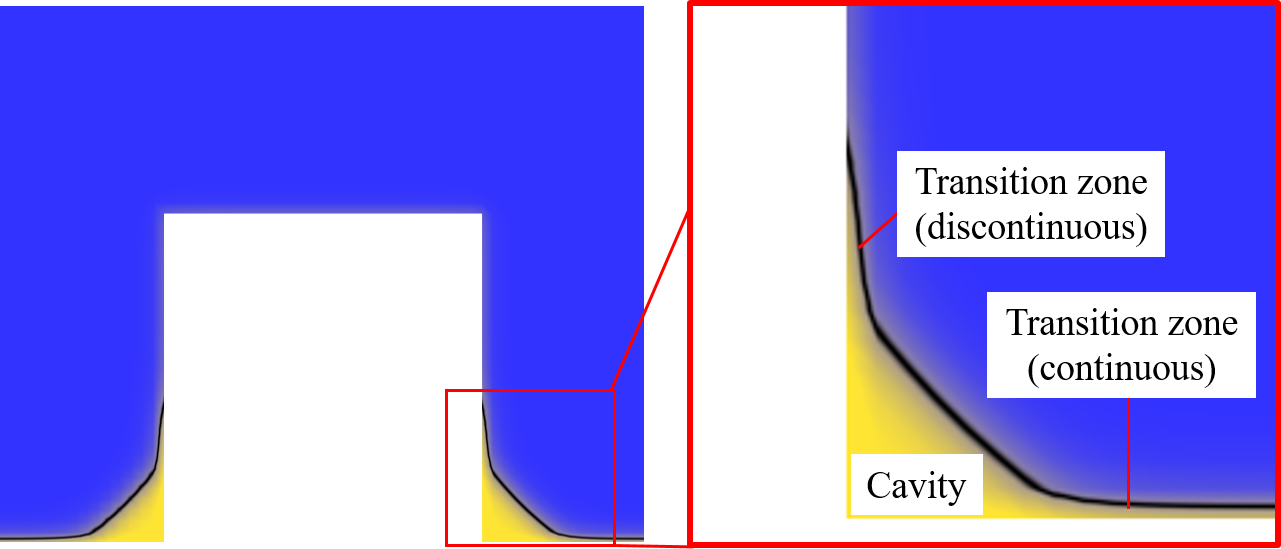}
\caption{Cross section of the phase field for a 3D post at $\theta_o = 150^\circ, \Delta P_r = 0.15$, showing a distorted vapour cavity (yellow) forming in the liquid phase (blue) at the base of the post (white).}
\label{fig:films}
\end{figure}

\section{Wetting films and line tension}

Fig. \ref{fig:films} shows a cross section of the cavity formed about a 3D post for $\theta_o = 150^\circ, \Delta P_r = 0.15$. Crucially, for $\theta_o$ greater than approximately $140^\circ$, the solid-fluid and liquid-vapour interface interactions become non-negligible and significantly change the shape of the liquid-vapour interface from a surface of constant mean curvature. These interactions are principally manifest in the transition zones \cite{Starov2007} with thickness on the order of the characteristic width of the liquid-vapour interface, $\epsilon$. Where they exist, the transition zones located on the sides of the pillar are typically observed to have a finite extent from the vapour cavity at all $\Delta P_r$ tested. The transition zones at the base of the system merge across the periodic boundaries, except where $\Delta P_r$ becomes large and the cavity size becomes small. Thus, the transition zones represent three-phase contact regions, leading to the occurrence of line tensions within the diffuse interface system.

\section{Pressure and the MEP}

\begin{figure}[!ht]
\centering
\includegraphics[width=0.6\textwidth]{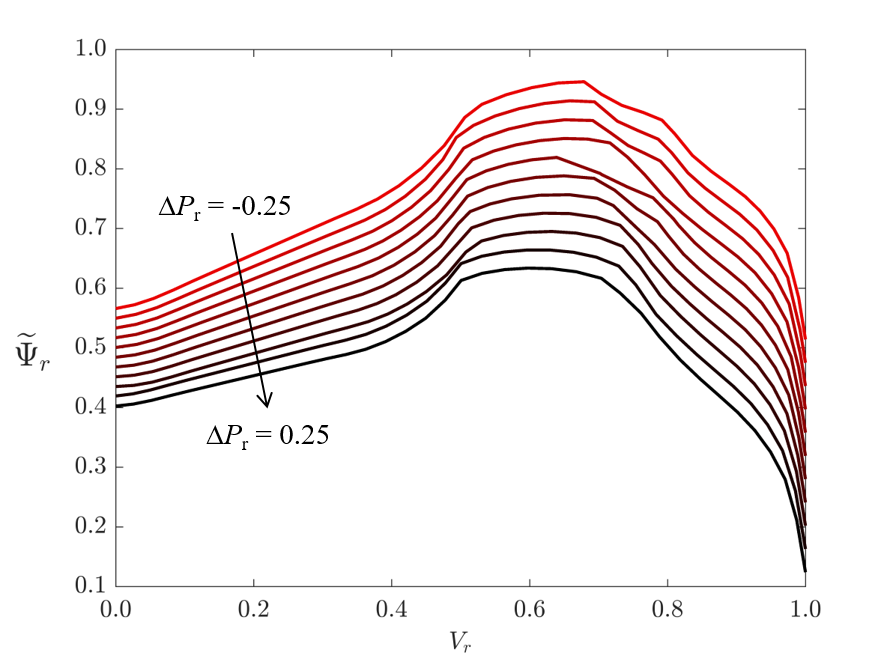}
\caption{Variation in the MEP from Suspended (top) to Collapsed for the short reentrant geometry in the reduced pressure range $-0.25 \leq \Delta P_r \leq  0.25$, $\theta_o=110^\circ$.}
\label{fig:pressures}
\end{figure}

For the tall and short posts and reentrant structures in both two and three dimensions, the MEPs were obtained for the wetting transition across the pressure range $-0.25 \Delta P_r < 0.25$, at $\theta_o = 110^\circ$. However, the principal effect of changing the pressure on the 	MEP was to shift the energy of each image in proportion to the volume of the liquid phase present. An example of this shifting is shown for the short 3D reentrant geometries in Fig. \ref{fig:pressures}.

\bibliography{References}